\documentstyle[11pt]{article}

\author{M.~A.~Nielsen \\
  Center for Quantum Computer Technology and Department of Physics, \\
  University of Queensland, Queensland 4072, Australia} 
\title{Introduction to quantum information theory}
\date{\today}

\begin{document}

\pagestyle{plain}
\pagenumbering{arabic}

\maketitle

\begin{quote}
  This is an expanded and revised text for a fifteen minute talk given
  at the University of Queensland Physics Camp, September 2000.  The
  focus is on the goals and motivations for studying quantum
  information theory, rather than on technical results.
\end{quote}

%
%
My name is Michael Nielsen, and I work at the University of Queensland
on {\em quantum information theory}, which is the subject of my talk
today.  It's part of a larger subject known as {\em quantum
  information science}, which is being investigated by many people at
the University of Queensland as part of the activities of the Center
for Quantum Computer Technology\footnote{A comprehensive introduction
  to quantum information science has been written by myself and Ike
  Chuang \cite{Nielsen00a}, and an excellent introduction by John
  Preskill is available free of charge on the web \cite{Preskill98c}.}.
Technical work on quantum information gets pretty mathematical pretty
quickly, but my talk today doesn't involve any equations.  Instead I
want to talk about the goals and motivation for quantum information
theory, and to try to convey some flavour of the subject.  Let me
start off by explaining what I mean by the term {\em quantum
  information theory}.

%
%
I would like each of you to imagine that you're a chess grandmaster
introduced to someone who claims to know all about chess.  You play a
game with this person and quickly discover that although they know all
the {\em rules} of chess, they have no idea of {\em how} to play a
good game.  They sacrifice their queen for a pawn, and lose a rook for
no apparent reason at all.  Naturally, you conclude that while they
know the rules of chess, this person does not in any sense {\em
  understand} chess.  That is, they don't know any of the high-level
principles, rules of thumb and heuristics which constitute a good
understanding of chess, and which are familiar to any master.

%
%
Humanity as a whole is in a similar position with respect to quantum
mechanics.  We've known all the basic rules of quantum mechanics for
quite some time, yet we have a quite limited understanding of those
rules and the higher-level principles they imply\footnote{This is, of
  course, not necessarily true of specific quantum phenomena like
  laser cooling or superconductivity, but these are extraordinarily
  specialized situations.  I'm talking about general properties of
  quantum mechanics.}.  As an example, consider that in
1982 \cite{Dieks82a,Wootters82a} it was discovered that the laws of
quantum mechanics prohibit the construction of a device which makes
perfect copies of unknown quantum states.  This ``no-cloning
principle'' is obviously an extemely important general heuristic
governing what is and is not possible in quantum mechanics, yet it was
only discovered 60 years after the basic rules of quantum mechanics
were found.  What other as-yet-unknown emergent properties are hidden
within the fundamental laws of quantum mechanics?

%
%
One way of answering the question ``What is quantum information
theory?'' is to say that it's the quest to obtain a set of higher
level principles and heuristics about quantum mechanics analogous to
those which a master chess player uses when playing chess.  This quest
for understanding is not dissimilar to (but goes beyond) the kind of
understanding that Dirac referred to when he said that ``I understand
what an equation means if I have a way of figuring out the
characteristics of a solution without actually solving
it.''\footnote{Quote taken from Feynman, Leighton and Sands
  \cite{Feynman65b}, page 2-1.}  The understanding we are pursuing in
quantum information theory exceeds even this, since we want to know
qualitatively what phenomena are {\em possible} within quantum
mechanics.

%
%
I've talked a little about one of the main goals of quantum
information theory, but haven't explained in concrete terms the
motivations one might have for wanting to pursue this goal.  In the
remainder of my talk I will describe two of the problems that
originally got me excited about quantum information theory, and which
continue to motivate much of my work today.

%
%
The first question is ``What does it mean to compute?''  To explain
what this question means we have to go backwards in time and review a
little history.  We'll start in 1936 with one of my all-time favourite
scientific papers, Alan Turing's paper \cite{Turing36a} on the
foundations of computer science.  Turing made three astonishing leaps
in this paper, of which only the first two are relevant to our story,
so I'll omit the third from my discussion
today\footnote{Unfortunately, in the interests of brevity I'm also
  omitting the role played by many other great researchers, such as
  G\"odel, Church, Tarski and Post.  See \cite{Hofstadter79a} for more
  on the history.}.

%
%
Turing's first innovation was to {\em mathematically define} the
process of computation.  Prior to Turing the notion of computation was
rather vague and ill-defined.  Turing introduced a mathematically
precise definition of computation, which made it amenable to study by
the powerful methods of modern mathematics, an innovation comparable
to the leap forward made by Galileo and Newton in bringing physics
into the realm of the mathematical sciences.  Turing's second great
innovation was to introduce the notion of a {\em universal computing
  device}.  That is, he had the idea that there might be a single,
simple computing device capable of simulating any other computing
device.  Although familiarity now makes this notion appear obvious,
{\em a priori} it is not remotely obvious that the Universe is such
that to analyse all possible computations we may restrict our
attention to a single type of computing device.

%
%
Turing's visionary work has some shortcomings.  In particular, the
fundamental thesis that his model of computation suffices to describe
all possible computations is open to attack.  Turing justified this
thesis, now known as the Church-Turing thesis, on a rather {\em ad
  hoc} basis, based on introspection and simple empirical arguments.
Attempts to confirm or refute the Church-Turing thesis continued for
50 years after Turing, without resulting in any major challenges to
the thesis, but leaving the thesis still on a rather {\em ad hoc}
footing.

%
%
Let's jump forward in time to 1985, when David Deutsch wrote a
terrific paper \cite{Deutsch85a} that represents the next great step
towards justification of the Church-Turing thesis.  Deutsch has many
good ideas in his paper, of which I'll describe three.  Deutsch's
first good idea was that it might be possible to {\em derive} the
Church-Turing thesis from the laws of physics.  Instead of having an
{\em ad hoc} basis, the thesis would then be on ground as firm as the
laws of physics themselves.  Deutsch's second idea was that it might
be possible to prove a particular device universal for computation,
starting from the laws of physics.  That is, a single universal
computing device would be capable of simulating any other physical
process!  Think about how remarkable this property would be: it would
mean that a single physical system was capable of simulating any other
physical system whatsoever; a relatively simple piece of the Universe
would in some sense contain all the rest!  {\em A priori} it need not
be true that the laws of physics allow such a device to exist, yet
Deutsch proposed that it might be so.

%
%
Deutsch's third innovation was to propose a candidate universal
computing device, the quantum computer.  He didn't actually prove that
his proposed device was universal for computation, but he did make
some progress in that direction.  It remains one of the most
interesting open problems in quantum information theory to determine
if the quantum computer is universal or not.  Indeed, it is
conceptually useful to divide work on the theory of quantum
computation into two categories: research into the capabilities
afforded by the standard model of quantum computation {\em a la}
Deutsch and modern variants, and research into the validity of the
model, and whether it might be possible to extend the model.  For
example, might it be that effects from general relativity, quantum
field theory or quantum gravity may be used to achieve computational
capabilities more powerful than in the standard model of quantum
computation?  Conversely, might attempts to find a universal computing
device produce insights into the problem of producing a quantum theory
of gravity?

%
%
We've looked at one problem motivating quantum information theory, the
problem of understanding what it means to compute.  I want to talk now
about a more specific problem motivating quantum information theory,
the problem of understanding the principles governing the behaviour of
quantum entanglement.  In case you haven't yet met entanglement in
your classes, I'll give a brief description that summarizes the
essence, without relying on technical definitions.  One way of
thinking of entanglement is as a type of {\em physical resource},
rather analogous to energy.  More concretely, entanglement is a joint
property of two (or more) physical systems; we say that these systems
are in an {\em entangled state}.  For our purposes, the precise
mathematical description of this state won't matter.  What does matter
is that such entangled states can have physical properties that can't
be explained within our conventional ``classical'' view of the world,
an insight we owe to John Bell's \cite{Bell64a,Peres93a} pioneering
work on the Bell inequalities --- an early example of work on quantum
information theory!  In recent times it has been discovered that
entangled states can be used to do all manner of surprising tasks,
such as quantum teleportation \cite{Bennett93a}, superdense
coding \cite{Bennett92c} and quantum cryptography \cite{Ekert91a}.

%
%
I said that entanglement is a physical resource, analogous to the
physical resource energy.  By this, I mean that entanglement has
properties which are {\em representation independent}.  Energy can be
given to us in many different forms --- chemical, nuclear, electrical,
and so on --- but from a fundamental point of view it does not so much
matter what form we receive the energy in; it is the quantity of
energy itself that is important.  In a similar sense, it doesn't
matter in what form we are given a quantity of entanglement ---
whether we are given entangled photons, entangled atoms, or whatever
--- it is the amount of entanglement we are endowed with that matters.
In any given physical situation we can be endowed with a quantity of
entanglement (or energy), and such an endowment gives us the ability
to perform tasks --- such as teleportation or superdense coding ---
that would otherwise be impossible, independent of the exact form of
the entanglement.  This idea, that entanglement is a new type of
physical resource, is not at all obvious, and is quite a recent idea,
given that we've known about entanglement since the 1930s.  A ghost of
the idea can be discerned quite far back in the literature, but I
think it was first made explicit in the mid-1990s, especially through
the pioneering work of Bennett, Divincenzo, Smolin and Wootters
\cite{Bennett96a}.  Since that time, a great deal of effort by many
people has been devoted to finding a set of high level principles
governing the behaviour of entanglement, much as in the 19th century
people discovered the laws of thermodynamics, which are a set of
high-level principles governing the behaviour of energy.  We'd like to
find fundamental principles governing the creation, manipulation and
observation of entanglement.  Are there laws governing the transport
properties of entanglement in physical systems?  Might there be
conservation laws or inequalities relating the transformation of
entanglement into other types of physical resource?

%
%
One first step along the way to developing a coherent set of
high-level principles governing the utilization of entanglement is the
development of quantitative measures of {\em how much} entanglement is
present in a given quantum state.  To finish up, I'd like to mention
just one application of this idea.  Perhaps the biggest open problem
in theoretical computer science at the dawn of the twenty-first
century is to show that $\mathbf{P} \neq \mathbf{NP}$.  Don't worry if
you've never heard of this problem before --- it's related to showing
that many important and common computational problems are in some
fundamental sense {\em difficult}.  One way of appreciating the
importance of this problem is that the Clay Mathematics Institute
(www.claymath.org) is offering seven ``Millenium Prizes'' of a million
US dollars each for the solution of seven major problems in
mathematics.  The problem $\mathbf{P} \neq \mathbf{NP}$ is one of
those problems.  The connection with the quantification of
entanglement is this: it has recently been shown \cite{Nielsen00e}
that if the amount of entanglement in some class of quantum states
exceeds a certain lower bound, then $\mathbf{P} \neq \mathbf{NP}$!
Thus, the quantification of entanglement has the potential to inform
not only our understanding of quantum mechanics, but also our
knowledge of other, apparently unrelated areas of science.

\section*{Acknowledgments}
My view of quantum information theory has developed through
conversations with many people, especially Dorit~Aharonov, Ike~Chuang,
Chris~Fuchs and Ben~Schumacher.  Dorit~Aharonov originated the elegant
and stimulating turn of phrase ``what does it mean to compute?''
Thanks also to Hideo~Mabuchi for a stimulating conversation about
analogies between entanglement and thermodynamics in early 1997.
Finally, thanks to Vladim\'{\i}r~Bu\v{z}ek for many interesting
discussions about quantum information theory as this talk was being
prepared, and to Damian~Pope and Mitch~Porter for helpful comments on
the manuscript.

\end{document}